# Introducing Schema Inference as a Scalable SQL Function [Extended Version]

Calvin Dani
Santa Clara University
Santa Clara, CA, USA
cdani@scu.edu

Shiva Jahangiri
Santa Clara University
Santa Clara, CA, USA
sjahangiri@scu.edu

Thomas Hütter[*]
Software Competence Center
Hagenberg
Hagenberg, Austria
thomas.huetter@scch.at

## Abstract

This paper introduces a novel approach to schema inference as an on-demand function integrated directly within a DBMS, targeting NoSQL databases where schema flexibility can create challenges. Unlike previous methods relying on external frameworks like Apache Spark, our solution enables schema inference as a SQL function, allowing users to infer schemas natively within the DBMS. Implemented in Apache AsterixDB, it performs schema discovery in two phases—local inference and global schema merging—leveraging internal resources for improved performance. Experiments with real-world datasets show up to a two orders of magnitude performance boost over external methods, enhancing usability and scalability.

## 1 Introduction

Semi-structured data has become more common due to web applications, IoT devices, social media, and cloud services generating flexible data formats such as JSON and XML. Unlike structured data with predefined schemas, this data is often stored in raw files or NoSQL DBMSs with flexible or missing schema definitions. While this independence from a rigid schema declaration offers greater flexibility for semi-structured data, it challenges users in understanding data structure and formulating accurate queries. "Schema Discovery" or "Schema Inference" (*SI*), automatically identifies the structure, attributes, and data types within semi-structured datasets. This process is crucial in environments with evolving data and inconsistent schema definitions, providing clarity and enabling effective data utilization despite its inherent variability.

Existing *SI* approaches [7, 17, 21, 22, 26, 27, 30, 33] are typically designed as standalone algorithms that depend on external data processing frameworks, such as Apache Spark [34]. While these frameworks are effective for schema discovery, they introduce several limitations when applied to schema discovery for data stored in a DBMS.

[*]This work was partially done while the author was at the University of Salzburg.

*(1) Efficiency:* Instead of performing *SI* within the DBMS, these methods use pipelines and connectors to link the DBMS to external frameworks that scan the data and infer the schema. This approach adds latency due to data movement and requires extra effort to set up a separate pipeline for schema inference. This complexity could be avoided if DBMSs natively support *SI* as a query feature.

*(2) Flexibility and Usability:* Current approaches predominantly require users to materialize the results of a query to infer the schema. This manual intervention reduces the level of automation for *SI*.

This study introduces a novel approach that integrates *SI* directly within a DBMS using a SQL-style function applicable to both base datasets and query results. Following the architecture and algorithms of parallel databases, our implementation of *SI* is horizontally scalable, minimizing data movement and leveraging DBMS metadata to address the efficiency and usability limitations of earlier approaches. Our empirical evaluations demonstrate up to a two orders of magnitude performance improvement over external methods, achieved through optimized internal processing and parallel execution.

The key contributions of this paper are:

- We present the first approach to natively support *SI* as an on-demand SQL function in DBMSs, demonstrated through an implementation in Apache AsterixDB [3]. Given its architectural similarities to systems like Google BigQuery [16], Couchbase [10], and CockroachDB [29], our approach can be applied to other parallel DBMSs.

- We perform a preliminary study to evaluate the performance of schema inference approaches using real datasets, unlike prior work focused mainly on algorithm design and schema quality.

The paper is structured as follows: Section 2 provides an overview of the inferred schema format, contemporary *SI* implementations and Apache AsterixDB. Section 3 details the implementation and structure of the *SI* operation. Section 4 evaluates the performance of *SI* within the DBMS and compares it to external approaches. Section 5 reviews related work while Section 6 concludes the paper and discusses future research directions.

## 2 Background

A database schema defines the structure of tables, relationships, and data types to ensure consistency and integrity. Traditional relational DBMSs enforce predefined schemas for consistency

```
{
  "productId": 1,
  "productName": "Ice sculpture",
  "price": 12.50,
  "tags": [ "cold", "ice" ]
}
```

**Listing 1: JSON Data Example for Product Catalog**

```
{
  "$schema": "https://json-schema.org/draft/2020-12
    ↪ /schema",
  "$id": "https://example.com/product.schema.json",
  "title": "Product",
  "type": "object",
  "properties": {
    "productId": {
      "type": "integer"
    },
    "productName": {
      "type": "string"
    },
    "price": {
      "type": "number"
    },
    "tags": {
      "type": "array",
      "items": {
        "type": "string"
      },
    }
  }
}
```

**Listing 2: JSON Schema for Product Catalogue**

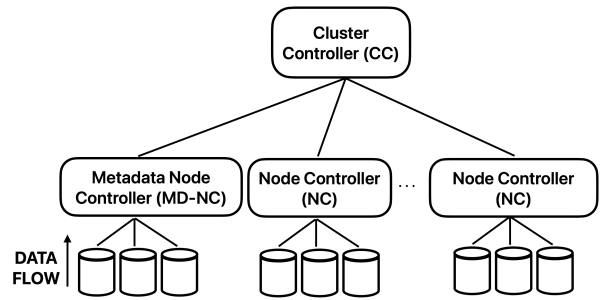

**Figure 1: Apache AsterixDB architecture**

but lack flexibility for evolving data. NoSQL databases such as DynamoDB [4], Couchbase [11], and MongoDB [23] offer more flexibility by inferring schemas at query time, but this can lead to data inconsistency and limited schema knowledge.

*SI* operations address these issues by automatically generating schemas, often in formats like JSON Schema, improving data model visibility and supporting dynamic data environments. Implementations of *SI* on platforms such as Apache Spark include works by Baazizi [7], Spoth [27], and Mior [22]. Our approach is inspired by Baazizi et al.[7] explore the trade-off between schema precision and succinctness, Spoth et al.[27] introduce a bi-clustering algorithm to identify entities for schema merging, while Mior [22] enriches schemas by maintaining monoid data structures (e.g., bloom filters) during merging, capturing additional metadata like minimum and maximum values for integers. These implementations require a data pipeline to Spark, introducing computational overhead.

### 2.1 JSON Schema Representation

Various representations for schemas of JSON documents have been proposed in literature, e.g., UML and class diagrams [17, 26]. In recent years, JSON Schema [24] became the de-facto standard format for schemas of JSON documents [22, 27, 33]. For example, Listing 2 shows the schema for the document in Listing 1 in JSON Schema format. The schema consists of an object that contains four properties (i.e., attributes). Next to basic functionality as shown Listing 2, JSON Schema allows for elaborated concepts such as bounding the values of integers, limiting the number of elements in an array, or referencing repeating sub-schemas, through the use of validation vocabulary defined by its standard [18].

### 2.2 Apache AsterixDB

Apache AsterixDB[2, 20] is a Big Data Management System (BDMS) designed to manage semi-structured data across multiple nodes in a shared-nothing cluster architecture. It uses AsterixDB Data Model (ADM) which is a super-set of JSON providing flexibility in record structure[3]. Apache AsterixDB natively supports spatial, temporal, and textual data types, comparable to advanced capabilities in systems like PostgreSQL and Oracle. AsterixDB uses SQL++ as its query language which is an extension of SQL designed for querying semi-structured data. SQL++ enables querying nested documents and arrays using path expressions, while distinguishing between null, missing, and undefined values—crucial in NoSQL environments. It offers enhanced support for nested objects and arrays, simplifying complex data navigation, while retaining SQL's familiar syntax for easy adaptation. AsterixDB's primary organizational unit is "dataverse" similar to databases in relational systems. Within a dataverse, users define "data types" to establish the structure of each record and create "datasets" which are collections of records that adhere to these predefined data types. Apache AsterixDB supports a wide range of data types to manage both structured and semi-structured data. These include primitive types such as integers, floats, booleans, and strings, along with complex types such as records (similar to JSON objects), arrays, and multisets for handling collections of values. AsterixDB supports both open and closed dataset types. Open datasets allow records to contain additional fields beyond those defined in the schema, offering flexibility for evolving data. Closed datasets, on the other hand, enforce a fixed structure, similar to relational databases. In closed datasets, schema information is stored once in the metadata, making them space-efficient. For open datasets, while predefined attributes are stored in the metadata, any extra fields unique to individual records are stored within the records themselves, which can result in higher storage overhead.

In Apache AsterixDB data can be ingested through loading, insertion queries, and accessed via queries that return results. As illustrated in Figure 1, the Cluster Controller (CC) serves as the logical entry point for user requests, while the Node Controllers and Metadata Node Controller (MD-NC) offer access to AsterixDB's metadata and leverage the aggregate processing power of the underlying shared-nothing cluster [3]. Apache AsterixDB stores data in partitions across NCs, distributing datasets for scalability and parallel processing. Each partition is managed by an NC, enabling concurrent query execution across multiple nodes. For multi-core processing, AsterixDB assigns a default number of working threads equal to the data partitions, allowing parallel processing. Users can adjust this parallelism to match the number of CPU cores. Working threads handle data processing, while separate I/O threads manage input/output tasks, ensuring efficient use of resources.

We selected Apache AsterixDB as the primary testbed for implementing and evaluating our proposed techniques for several reasons. First, as an open-source platform, it enables us to share our techniques and evaluation results with the broader community. Second, AsterixDB is a parallel big data management system specifically designed for managing and processing large volumes of semi-structured data using a declarative language. Its support for defining open datasets was crucial for making schema inference useful. Third, the structural, parallel, scalable, and distributed design similarities between AsterixDB and other NoSQL and NewSQL systems, such as Google BigQuery [16], Couchbase [10], and CockroachDB [29], demonstrate that the techniques and findings from AsterixDB are broadly applicable across a variety of database systems. These systems share architectural principles like horizontal scaling, distributed storage, and parallel query execution, making AsterixDB's approaches to handling semi-structured data, parallelism, and scalability relevant for improving performance in other distributed and scalable databases.

## 3  *SI* as a SQL Function: Design

The implementation of *SI* as a SQL function offers several advantages, including user familiarity with SQL functions and the ability to infer the schema of both query results and base datasets. This approach leverages the existing database architecture for parallel record processing and utilizes metadata efficiently. We have implemented two types of *SI* operations in Apache AsterixDB: Open-*SI* and Closed-*SI*. The Closed-*SI* operation generates a JSON schema from the metadata of a dataset defined as Closed (structured data), while the Open-*SI* derives the schema by scanning the records of a dataset defined as Open (semi-structured data), as described in further detail below. We first explain the rationale and operation of the Open-*SI* as an aggregate function, then introduce the Schema Intermediate Structure (*SIS*), a data structure used to model records and their schema, and finally discuss how *SIS* is employed and implemented within the SI aggregate function.

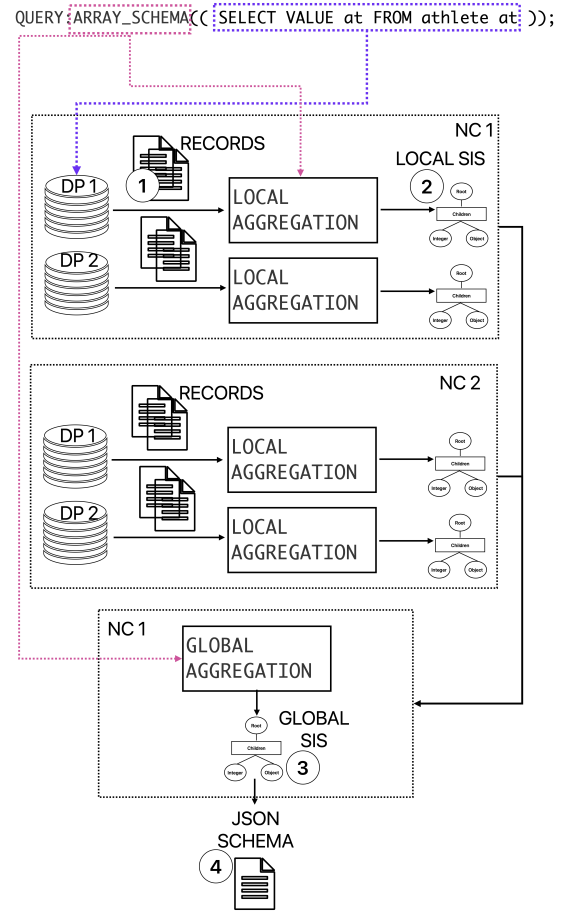

Figure 2: *SI* Aggregate Query Execution in a Database

### 3.1  Leveraging Aggregate Functions for *SI*

We developed two *SI* approaches in AsterixDB: Open-*SI* for semi-structured datasets and Closed-*SI* for structured datasets. Both approaches allow users to infer schemas easily using simple SQL queries, such as the following for Open-*SI*:

ARRAY_SCHEMA((SELECT VALUE athl FROM athlete athl));

Open-*SI* dynamically scans query results to generate schemas for flexible datasets, while Closed-*SI* retrieves predefined schemas directly from metadata for structured datasets, skipping the need to scan records. Both methods can be adapted for other DBMSs with adjustments.

Open-*SI* functions similarly to the AVERAGE aggregate function in parallel DBMSs, as seen in Fig. 2. It starts by executing the input query and infers the schema as data is pipelined to the *SI* operator labeled as 1. Local schemas are created in each data partition labeled as 2 and then sent to a central NC, where they are merged into a global schema labeled as 3. The result is then converted into a JSON Schema labeled as 4 for external use.

Closed-*SI* functions by fetching the defined schema for the dataset from the metadata node of the database and then converting it to a JSON Schema. A simple SQL query for Closed-*SI* is described below:

```
DECLARED_SCHEMA((athlete));
```

We use a Schema Intermediate Structure (*SIS*) to internally represent schemas, inspired by Bazizi's work [7], which will be further explained in the next subsection.

### 3.2 Schema Intermediate Structure

The Schema Intermediate Structure (*SIS*), a tree-based structure capable of capturing hierarchical organization, is used to represent structure of records and be transformed into a JSON Schema. The structure of *SIS* builds on previous work of [2]. Each *SIS* node represents a component of the *ADM*.

The nodes of the *SIS* structure include Primitive, Array, Multiset, Object, and Union. We will describe how these nodes aim to model the ADM. Nodes can be categorized into three distinct types:

(1) Primitive Nodes: These represent fields with existing predicates on the record, such as strings and integers. Primitive Nodes encapsulate basic data types that do not require further decomposition.
(2) Composite Nodes: These nodes represent fields that contain nested levels of data, such as objects and arrays. Composite Nodes represents complex data structures with hierarchical relationships.
(3) Union Nodes: These nodes define fields that can have multiple types. Union Nodes achieve this flexibility by allowing a single field to be associated with two or more data types, enabling the merging of contradictory fields into one when schemas differ between two records or between two schemas.

An example of SIS is illustrated in Fig. 3 labeled as "C" and "D" and is further explained in the next subsection.

### 3.3 Schema Inference Algorithm

The Open-*SI* mechanism leverages the SIS tree and Apache AsterixDB's aggregate framework to model each record's schema and generate a JSON Schema. Fig. 3 shows how Primitive ("Id") and Composite ("Friend") nodes are integrated into a *SIS* tree. Composite nodes recursively add all descendants before the node itself, while Primitive nodes are added directly.

The process is outlined in Algorithms 1 and 2. For fields with mixed types, such as "Peeve" which contains both Integer and String values, a Union node is created to resolve type conflicts following Algorithm 3. This ensures accommodation of diverse data types within a single field. The Global *SIS* is built by merging each Local *SIS* rather than individual records, following the same process as the Local phase using Algorithms 4, 2, and 3. For Union nodes, the Global node combines all field types from the merging Union nodes.

---

**Algorithm 1** computeSIS - Local Aggregation Phase

**Input:** Records $R$
**Output:** Schema Intermediate Structure $SIS$
$SIS \leftarrow$ SIS
**for** $record$ **in** $R$ **do**
  **for** $field$ **in** $record$ **do**
    **if** $field.name$ in $SIS$ **then**
      $node \leftarrow SIS[field.name]$
      **if** $node.type$ **is not** $field.type$ **then**
        $SIS[field.name] \leftarrow$
          createUnionNode($node.type, field$)
    **else**
      $SIS[field.name] \leftarrow$ createNode($field$)
**return** $SIS$

---

**Algorithm 2** createNode

**Input:** Field $f$
**Output:** SIS node $node$
**if** $f$ **has** children **then**
  $node \leftarrow$ node($f.type$)
  **for** $nestedField$ **in** $f.children$ **do**
    $node.children \leftarrow node.children \cup$
      (createNode($nestedField$))
**else**
  $node \leftarrow$ node($f.type$)
**return** $node$

---

**Algorithm 3** createUnionNode

**Input:** SIS node $node$ and field $f$
**Output:** Union node $uNode$
$uNode \leftarrow$ union(($\varnothing$))
**if** $node.type$ **is not** $f.type$ **then**
  **if** $node.type$ **is** 'Union' **then**
    $node.child \leftarrow node.child \cup$ createNode($f.type$)
    $uNode \leftarrow node$
  **else**
    $uNode \leftarrow uNode.child \cup$ createNode($f.type$)
**return** $uNode$

---

**Algorithm 4** mergeSIS - Global Aggregation Phase

**Input:** Collections of locally aggregated SISs $C$
**Output:** Globally merged SIS $globalSIS$
**for** $SIS$ **in** $C$ **do**
  **for** $node$ in $SIS$ in DFS traversal **do**
    **if** $node.field.name$ in $globalSIS$ **then**
      $existingNode \leftarrow SIS[node.field.name]$
      **if** $node.field.type$ **is** 'Union' **or**
        $existingNode.field.type$ **is** 'Union' **then**
▷ mergeUnionNodes combines two nodes fields, returning a single union node
        mergeUnionNodes($node, existingNode$)
      **else if** $node.type$ **is not** $existingNode.type$ **then**
        createUnionNode($node,$
          $existingNode.field$)
      **else**
        $SIS[field.name] \leftarrow$
          createNode($node.field$)
**return** $globalSIS$

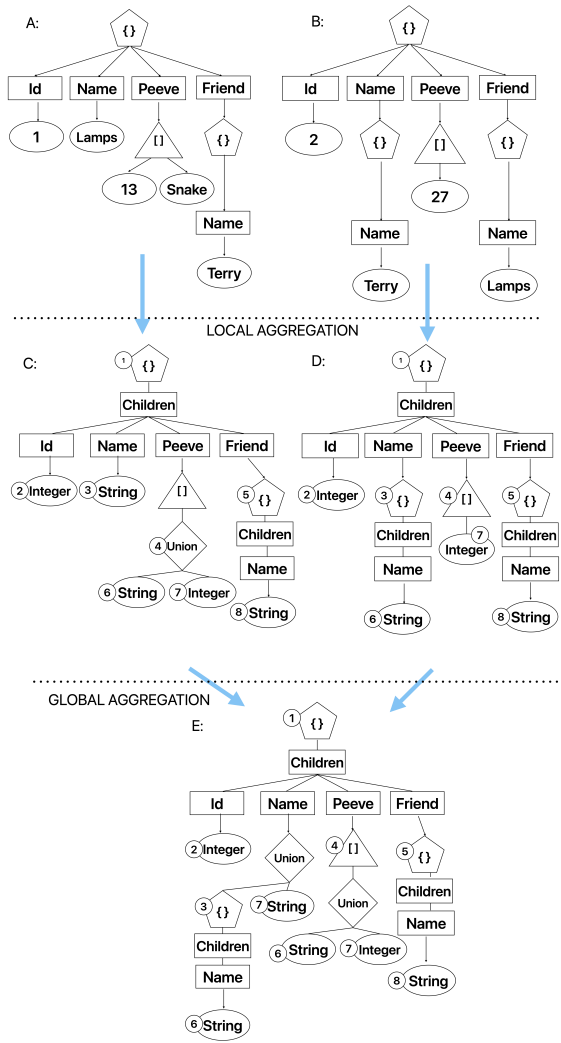

Figure 3: Visualization of JSON Trees labeled "A" and "B" and *SIS* labeled "C", "D" and "E" in *SI* Operation. Object nodes are represented as pentagons with the symbol "{ }", array nodes as triangles with the symbol "[ ]", keys as rectangles, and primitives as ellipses. The figure illustrates the *SI* operation in a database.

In the Closed-*SI* function, the *SIS* is directly generated from the predefined schema stored in the metadata, as it can be mapped to *SIS* nodes.

Once the *SIS* tree is built, either through Open- or Closed-*SI*, it is converted into JSON Schema via DFS traversal of the *SIS* tree, applying the JSON Schema standard [19]. THe steps include:

(1) Primitive nodes creates an object with the "type" keyword, containing its data type.
(2) Composite nodes are resolved and attached to the parent using keywords like "properties" for objects or "items" for arrays.
(3) Union nodes are handled by appending all fields under the "oneOf" keyword to define multiple data types.

The creation of a JSON Schema from an SIS tree can be illustrated through a simple example. Consider a tree representing a person's basic information. As we traverse the tree, we encounter a Composite node for the person object. This node is represented using the "type" keyword with the value "object" and a "properties" keyword to define its structure. Within the person object, we find Primitive nodes for name and age. The name node is assigned the "type" keyword with the value "string", while the age node uses "type" with "number". If the person object includes a Union node for contact information, allowing either an email or phone number, we would use the "oneOf" keyword to specify these alternatives. Each option within the Union would be defined as a separate object with its own "type" and relevant properties. This process of applying specific keywords based on node types continues until the entire SIS tree has been converted into a comprehensive JSON Schema.

The resulting JSON Schema is then sent to the user executing the *SI* operation. By integrating *SI* with existing database operations, we simplify the process of understanding complex datasets and handling unstructured data. Our Open- and Closed-*SI* implementations are available on GitHub [8].

## 4 Experiments

This section presents an experimental analysis of our proposed SI operation in Apache AsterixDB, followed by a comparison with related works. We evaluate the SI function's scalability and performance using Speed-Up and Scale-Up experiments, then compare Apache AsterixDB's performance with approaches by Baazizi et al. (2017) [7], Spoth et al. (2021) [27], and Mior et al. (2023) [22].

### 4.1 Configurations and Settings

We designed and conducted our experiments using a cluster of Intel NUC machines. Each node in the cluster features an 11th Gen Intel(R) Core(TM) i7-1165G7 CPU, operating at 2.80GHz. This single-socket processor has 4 physical cores, each supporting 2 threads via Intel's Hyper-Threading technology, providing a total of 8 logical CPUs per node. Each node is equipped with 64 GB of RAM and 2TB of SSD storage. The nodes are interconnected using a 2.5 Gbps Ethernet network, with a connection latency of less than 1 millisecond.

In our experimental setup for the investigation of schema inference implementation within Apache AsterixDB, we deployed Apache AsterixDB on the cluster consisting of a CC (Cluster Controller) and one or more NCs (Node Controllers) with each having one or more I/O device mount points. The specifics of these configurations for each experiment of Apache

| Expr. ID | # NCs | # DPs | Dataset Size |
|---|---|---|---|
| 1 | [1,2,4] | [1,2,4,8] | 220 GB |
| 2 | [1,2,4] | 1 | [55,110,220] GB |
| 3 | 2 | 8 | Refer to Table 2 |

Table 1: AsterixDB's Configurations

AsterixDB is detailed in Table 1 where "NCs" refer to individual Intel NUC machine and "DPs" refer to the data partition configuration.

To ensure consistency while investigating the schema inference implementation configuration are mirrored to be as similar as possible between Apache Spark and Apache AsterixDB platforms. The configuration for Apache AsterixDB included 2 Node Controllers (NCs) with 8 partitions per NC. Apache Spark was configured with 2 worker nodes, utilizing Hadoop YARN as the cluster manager. To facilitate data access across worker nodes, we employed Hadoop Distributed File System (HDFS) as the distributed file system for data storage. The HDFS configuration comprised 2 nodes, co-located on Intel NUC machines with the worker nodes. We utilized Hadoop Disk Balancer to evenly distribute data blocks across the node configuration. To maximize resource utilization, each schema inference implementation was submitted to Apache Spark as a Spark job with a configuration of 8 cores for the executor in the worker node, while Apache AsterixDB's 8 data partition were selected to fully leverage the available CPU cores and threads of the Intel NUC machines. Both Apache Spark and Apache AsterixDB were allocated 3 GB of JVM memory. To constrain memory usage to the configured limits, we blocked 55 GB out of the 64 GB of main memory. The remaining memory was designated for the experimental setup, encompassing the memory allocated to Spark and Apache AsterixDB during program execution, as well as operating system requirements.

The queries for each experiment runs for 11 iterations, where we report the average execution time of the last 10 iterations. To mitigate the impacts of file system cache, a dataset 5 times larger than the total system's memory is scanned after each query execution. For the evaluation of Spark implementations, we systematically dropped the Linux file system cache after each Spark job execution for eliminating potential performance variations due to caching. Additionally, Resilient Distributed Datasets (RDDs) were not configured for caching in memory or external distributed file systems across all the implementations [5]. Next, we will discuss and analyze the results of each experiment individually.

### 4.2 Datasets

In our experimental design, we prioritized real-world datasets for a realistic evaluation of our *SI* approach. For Expr. 1 and 2, we used a Twitter dataset with 17 million tweets totaling 220 GB. The datasets used for Expr. 3 included GitHub, Yelp, and Pharma datasets [25, 28, 32], consistent with prior studies [22, 27]. The datasets utilized for Expr. 3 are listed in Table 2, with the prefix "Y_" denoting the "Yelp" datasets. This table highlights diverse dataset properties, including the number of records, nodes, objects, keys, and max depth for each dataset, with the first element in an array denoting the minimum number of the said property and the second element denoting the maximum of the said property. $\mu$ denotes the average.

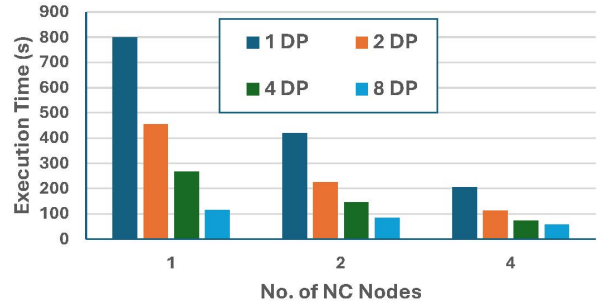

Figure 4: Speed-Up Evaluation of AsterixDB's Open-*SI*

### 4.3 Expr. 1: Open-*SI* Speed-Up Evaluation

This experiment evaluates the performance improvements of our Open-*SI* function under controlled conditions, maintaining constant dataset size while increasing the number of DPs leads to increasing number of I/O and computation threads up-to the number of CPU cores, hence increasing computation resources. This experiments aims to evaluate the performance gain of the system when dataset size remains the same, but the computation power of the system is increased, following Speed-Up definition. For each cluster configuration, we modify the number of DPs from 1 to 2,4, and 8. Each cluster configuration will have 1, 2 and 4 NCs.

Fig. 4 shows that increasing DPs reduces query runtime across all NC configurations. Doubling DPs in each setting reduces execution time to slightly more than half, which is closely inline with expected results of a Speed-Up evaluation. The deviation from perfect halving is due to the global phase, a sequential step that doesn't gain from parallel processing.

### 4.4 Expr. 2: Open-SI Scale-Up Evaluation

In this experiment, we evaluate the scalability of Apache AsterixDB Open-*SI* approach by proportionally increasing both the data size and the number of NCs, successively doubling them. Starting with 57 GB of tweets on 1 NC and consistently keeping the number of DP equal to 1 to avoid inter-core interference, we scaled up to 2 and 4 NCs. As Fig. 5 shows, our Open-*SI* approach can effectively scale with the increased number of worker nodes and data size, providing a robust and consistent performance inline with expected result from a Scale-Up test. The slight deviation from a perfect straight line in Fig. 5 is due to minor skew in data partitioning, causing one NC to handle slightly more records than the others.

### 4.5 Expr. 3: Comparison with Other Approaches

We assessed the performance of AsterixDB's Open- and Closed-*SI* functions against other schema inference approaches. AsterixDB was configured with 2 NCs and 8 data partitions per

Table 2: Natural Datasets' and Their Descriptive Statistical Information used in Expr. 3, where $\mu$ is the "Average"

| Statistics | Y_Checkin | Y_Business | Y_Tip | Pharma | Y_User | Y_Review | GitHub |
|---|---|---|---|---|---|---|---|
| Dataset Size | 274M | 114M | 173M | 157M | 3.2G | 5G | 51G |
| # Recs | 131930 | 150346 | 908915 | 239930 | 1987897 | 6990280 | 10649574 |
| # Nodes/Recs | [5, 5] | [29, 109], $\mu = 55.7$ | [11, 11] | [23, 1001], $\mu = 61.2$ | [45, 45] | [19, 19] | [45, 591273], $\mu = 322.67$ |
| # Objects | [1, 1] | [1, 3], $\mu = 2.7$ | [1, 1] | [3, 3] | [1, 1] | [1, 1] | [1, 17917], $\mu = 14.11$ |
| # Keys | [2, 2] | [14, 54], $\mu = 27.3$ | [5, 5] | [11, 500], $\mu = 30.1$ | [22, 22] | [9, 9] | [22, 286678], $\mu = 153.78$ |
| Max Depth | [2, 2] | [2, 4], $\mu = 3.9$ | [2, 2] | [2, 5], $\mu = 3.2$ | [2, 2] | [2, 2] | [2, 5], $\mu = 3.18$ |

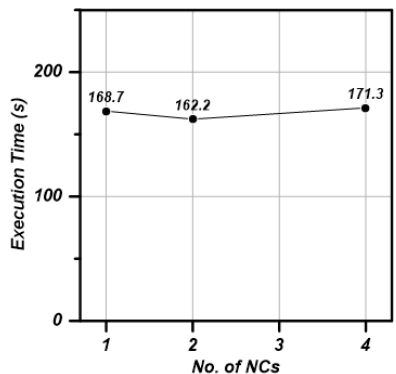

Figure 5: Scale-Up Evaluation of AsterixDB's Open-*SI*

NC, while Spark used two worker nodes and HDFS for data storage. Both platforms were fully optimized to utilize 8 CPU cores.

Our method was compared against Spark-based implementations by Bazizi [7], Spoth [27], and Mior's "Min" and "Simple" modes [22] (excluding the "All" mode due to configuration limits).

The results showed significant differences in execution time. Bazizi's method, which parses JSON character by character, was slower than AsterixDB's native processing. Mior's "Min" mode performed similarly to Bazizi's, while Spoth's was slower due to the bi-clustering algorithm. "Simple" mode took the longest time across all datasets due to expensive Monoid calculations, as illustrated in Fig. 6.

Execution times increased with dataset size and complexity. The Pharma dataset, with its large number of nodes and keys per record, caused longer processing times despite fewer records. AsterixDB's Open- and Closed-*SI* demonstrated the fastest performance, outperforming Mior by two orders of magnitude on Pharma and being an order of magnitude faster than Bazizi on Yelp Checkin. Closed-*SI* consistently finished in 0.03 seconds across all datasets by generating schemas directly from metadata without needing to scan records.

### 4.6 Results and Findings

The Speed-Up and Scale-Up experiments highlighted our *SI*'s ability to scale efficiently within and across NCs, taking full advantage of Apache AsterixDB's parallel architecture.

AsterixDB's Open-*SI* stood out, outperforming other methods by using the ADM format [6], which avoids the costly parsing Spark implementations require. Both Open- and Closed-*SI* maintained strong, consistent performance, unlike Spoth's bi-clustering approach and Mior's Monoid-based method, which suffer from extra overhead. Closed-*SI* further benefits by directly pulling schemas from metadata, skipping the need to scan all records, a clear advantage over other approaches.

In contrast, external methods like Spoth and Mior introduce extra overhead with their reliance on Spark pipelines and pre-processing steps, such as requiring newline-delimited JSON or additional work for JSON Schema generation.

## 5 Related Work

In literature, numerous approaches for JSON schema inference have been presented. Commonly, these algorithms infer a schema for each individual document in a MapReduce fashion and merge them hierarchically until a final schema is computed. Early approaches by Izquierdo and Cabot [17] and Sevilla Ruiz et al. [26] suggest a set of heuristic rules that allow to identify different entities and references within the data. In order to distinguish optional and required attributes, Klettke et al. [21] store meta-data in the form of a Structural Identification Graph. Baazizi et al. [7] expand on the work of Collazo et al. [9] by discussing the trade-off between the precision and the succinctness of a schema. They develop two extreme merging strategies: (1) L-reduce only merges schemas with the same labels, which produces large but precise schemas. (2) K-reduce rigorously merges all schemas, which results in a concise schema with (probably) many optional attributes. Spoth et al. [27] presents a more sophisticated merging strategy that involves a bi-clustering algorithm to identify entities. To mitigate the runtime increase, they experimentally evaluate the precision and recall of their approach when sampling the input data. The approach by Mior [22] extracts and maintains monoid data structures (e.g., bloom filters) during the merging process in order to enrich the schema with additional information. For example, the schema might contains the minimum and maximum values for integer values. Yun et al. [33] develop a bottom-up strategy that adopts the Minimum Description Length (MDL) from XML schema inference [15] to navigate through the search space. Further algorithms have been proposed by Wischenbart et al. [31] and Frozza et al. [14].

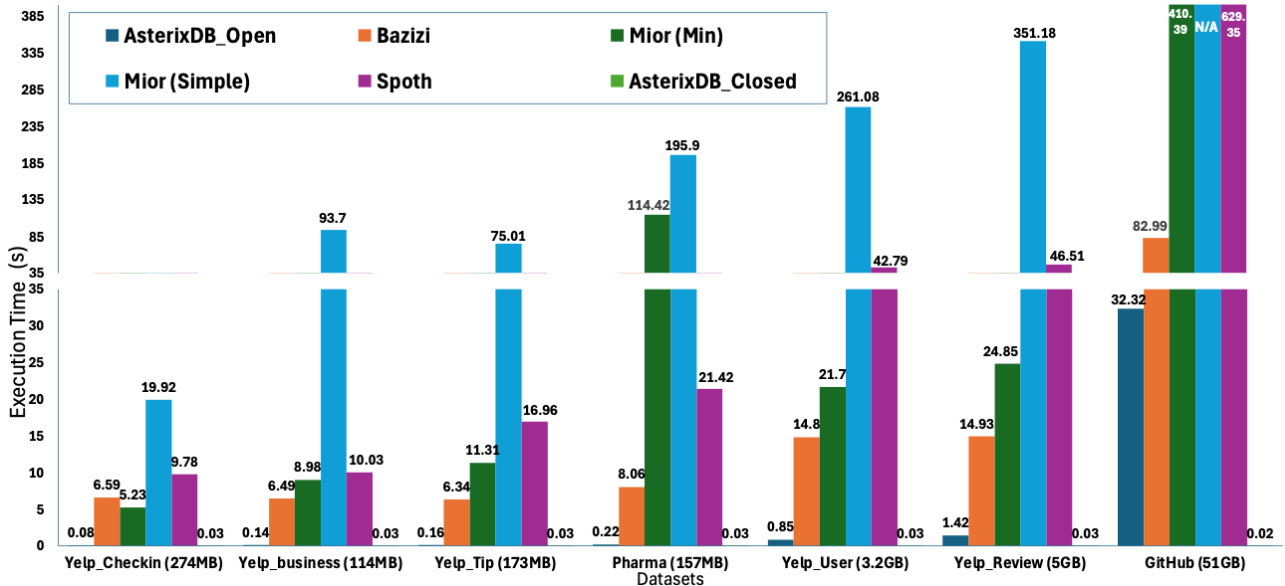

Figure 6: Expr. 3: *SI* operation Runtime for Datasets Referenced in Section 4.2

Compared to our solution, these approaches have been implemented and analyzed in a standalone fashion. Consequently, users have to manually extract and pre-process the data in order to compute a schema. By integrating JSON schema inference algorithms into databases, we enable users to formulate declarative queries to natively compute a schema within the system itself.

In a system context, JSON schemas have been used in various scenarios. Durner et al. [13] aim to store JSON documents in traditional databases by relationalizing the data. The presented approach, called JSON Tiles, partitions the data into groups with similar attributes and extracts a local schema for each group. Similarly, DiScala and Abadi [12] develop an unsupervised machine learning algorithm to automatically transforms JSON data into traditional relations. To reduce the storage overhead in Apache AsterixDB, Alkowaileet et al. [1] introduce a tuple compactor framework that infers a schema from semi-structured data during ingestion. Instead of using JSON schemas to improve a system's performance, we aim to facilitate the usage of JSON schema inference algorithms for users by providing native support within a database.

Wang et al. [30] introduce a framework supports inferring, managing, and querying of schemas from JSON data within a repository. Instead of developing system that is limited to this use case, we integrate our solution in a full-fledged database system.

There are various formats to describe the schemas of JSON collections. Initially, class and UML diagrams have been used [17, 26]. Baazizi et al. [7] suggest a well-defined type language for schemas along with a set of syntactic congruence rules. The schemas inferred by our solution are formatted in JSON Schema [24], which became the de-facto standard format in recent years.

## 6 Conclusion and Future Work

This paper presented a scalable *SI* approach, implemented in Apache AsterixDB, that leverages parallel database architecture for improved performance and scalability. A key contribution is offering *SI* approach as a SQL-based function, enabling users to explore data structures directly within the DBMS. Our experiments showed that Open- and Closed-*SI* approaches outperforms external methods such as Apache Spark by up to two orders of magnitude in some cases, due to parallel processing and avoiding data transfers.

For future work, we will explore using internal DBMS knowledge and metadata to reduce computation, including sampling based on indexes and leveraging predefined schemas to avoid record scanning. We also plan to combine metadata with the current method for datasets that are flexible yet share common schemas and enable statistics collection for attributes such as min/max values and null percentages.

## Acknowledgement

Special thanks to Michael J. Carey and Wail Y. Alkowaileet for their invaluable feedback and to the Apache AsterixDB team for their continuous support and recommendations.